\documentclass[preprint,superscriptaddress,showpacs,preprintnumbers,amsmath,amssymb]{revtex4}

\usepackage{graphicx}
\usepackage{dcolumn}
\usepackage{bm}

\def\lesssim{\ \raise.3ex\hbox{$<$}\kern-0.8em\lower.7ex\hbox{$\sim$}\ }
\def\gesim{\ \raise.3ex\hbox{$>$}\kern-0.8em\lower.7ex\hbox{$\sim$}\ }

\def\rnum#1{\expandafter{\romannumeral #1}} 
\def\Rnum#1{\uppercase\expandafter{\romannumeral #1}} 
\begin{document}
\title{Low-dimensional pairing fluctuations and pseudogapped photoemission spectrum in a trapped two-dimensional Fermi gas}
\author{Ryota Watanabe}
\affiliation{Faculty of Science and Technology, Keio University,
3-14-1 Hiyoshi, Kohoku-ku, Yokohama 223-8522, Japan}

\author{Shunji Tsuchiya}
\affiliation{Department of Physics, Faculty of Science, Tokyo University
of Science, 1-3 Kagurazaka, Shinjuku-ku, Tokyo 162-8601, Japan}

\author{Yoji Ohashi}
\affiliation{Faculty of Science and Technology, Keio University,
3-14-1 Hiyoshi, Kohoku-ku, Yokohama 223-8522, Japan}

\date{\today}       
\begin{abstract}
We investigate strong-coupling properties of a trapped two-dimensional normal Fermi gas. Within the framework of a combined $T$-matrix theory with the local density approximation, we calculate the local density of states, as well as the photoemission spectrum, to see how two-dimensional pairing fluctuations affect these single-particle quantities. In the BCS (Bardeen-Cooper-Schrieffer)-BEC (Bose-Einstein condensation) crossover region, we show that the local density of states exhibits a dip structure in the trap center, which is more remarkable than the three-dimensional case. This pseudogap phenomenon is found to naturally lead to a double peak structure in the photoemission spectrum. The peak-to-peak energy of the spectrum at $p=0$ agrees well with the recent experiment on a two-dimensional $^{40}{\rm K}$ Fermi gas [M. Feld, \textit{et al}., Nature \textbf{480}, 75 (2011)]. Since pairing fluctuations are sensitive to the dimensionality of a system, our results would be useful for the study of many-body physics in the BCS-BEC crossover regime of a two-dimensional Fermi gas.
\end{abstract}
\maketitle

\section{Introduction}
The advantage of ultracold Fermi gases is the existence of highly tunable physical parameters\cite{Giorgini,Bloch,Ketterle}. A tunable pairing interaction associated with a Feshbach resonance\cite{Timmermans,Holland,Chin} enables us to study Fermi superfluids from the weak-coupling BCS regime to the strong-coupling BEC limit in a unified manner\cite{EAGLES,LEGGETT,NSR,SADEMELO,RANDERIA_BEC,Ohashi,Jin,Zwierlein,Kinast,Bartenstein}. The intermediate coupling regime (which is also referred to as the BCS-BEC crossover region in the literature) is useful for the study of strong-coupling physics. Since correlation effects are important in high-$T_{\rm c}$ cuprates\cite{Damascelli,Yanase,Lee,Fischer}, the BCS-BEC crossover physics in ultracold Fermi gases would be also useful for the study of this interacting electron system.
\par
In addition to the tunable interaction, one can also adjust the atomic motion by using an optical lattice\cite{Giorgini,Bloch,Ketterle}. For example, when a Fermi gas is loaded on a one-dimensional optical lattice, the particle motion in the the lattice direction is strongly suppressed for a very high lattice potential, which effectively realizes a two-dimensional gas. Thus, an optical lattice can be used to change the system dimension\cite{KOHL,FROHLICH,SOMMER}.
\par
Using this optical lattice technique, Feld and co-workers\cite{KOHL} have recently done the photoemission-type experiment on a two-dimensional $^{40}$K Fermi gas. In the BCS-BEC crossover region, they found that the photoemission spectrum exhibits a double peak structure (although the system is in the normal state). Although a similar anomaly has also been observed in a three-dimensional $^{40}$K Fermi gas\cite{STEWART,GAEBLER}, the double peak structure observed in the former low-dimensional system is more remarkable than the latter, which indicates that the low-dimensionality enhances this anomaly. Indeed, evolution of a double-peak structure in the rf (radio frequency)-spectrum has also been observed in a $^6$Li Fermi gas\cite{SOMMER}, when the system dimension is changed from the three- to two-dimension by using an optical lattice.
\par
In the case of a three-dimensional Fermi gas, it has been theoretically shown that the anomaly seen in the photoemission spectrum can be explained as a pseudogap phenomenon originating from strong pairing fluctuations\cite{LEVIN,TSUCHIYA1,TSUCHIYA2,TSUCHIYA3,WATANABE1,WATANABE2,HU,BULGAC1,BULGAC2}. In this regard, we note that pairing fluctuations are stronger in a lower dimensional system. Thus, although the existence of the pseudogap is still controversial in ultracold Fermi gases\cite{Nascimbene}, if the pseudogap scenario proposed in the three-dimensional case is correct, the anomalous photoemission spectrum observed in a two-dimensional $^{40}$K Fermi gas is also expected to originate from pairing fluctuations that are enhanced by the low-dimensionality of the system.  
\par
In this paper, we investigate single-particle properties of a two-dimensional Fermi gas. Including two-dimensional pairing fluctuations within a strong-coupling $T$-matrix approximation\cite{TSUCHIYA1,PERALI}, as well as effects of a harmonic trap within the local density approximation (LDA)\cite{Pethick}, we calculate the local density of states in the normal state. We also deal with the photoemission spectrum, to see if the observed double peak structure can be understood as a two-dimensional pseudogap phenomenon. We briefly note that the possibility of the Berezinskii-Kosterlitz-Thouless (BKT) transition\cite{Berezinskii,Kosterlitz1,Kosterlitz2} in a two-dimensional Fermi gas have recently been discussed by many researchers\cite{Randeria,Melo1,Melo2,Zhang,Tempere}. Pseudogap physics in this system has also been studied in Refs.\cite{Pietila,Tempere}.
\par
This paper is organized as follows. In Sec. II, we explain our formulation. In Sec. III, we examine the local density of states. Here, we show how two-dimensional pairing fluctuations affect this single-particle quantity, leading to the pseudogap phenomenon. In Sec. IV, we discuss the photoemission spectrum. We compare our results with the recent experiment on a $^{40}$K Fermi gas done by Feld and coworkers\cite{KOHL}. Throughout this paper, we set $\hbar=k_{\rm B}=1$, and the system area $S$ is taken to be unity, for simplicity.
\par

\section{Formulation}
We consider a two-dimensional Fermi gas with two atomic hyperfine states, described by the BCS Hamiltonian,
\begin{equation}
H=\sum_{{\bm p},\sigma}\xi_{\bm p}c_{{\bm p}\sigma}^\dagger c_{{\bm p}\sigma}-U\sum_{\bm {p,p^\prime,q}}c_{{\bm {p+q/2}}\uparrow}^\dagger c_{{\bm {-p+q/2}}\downarrow}^\dagger c_{{\bm {-p^\prime+q/2}}\downarrow}c_{{\bm {p^\prime+q/2}}\uparrow}.
\label{Hamiltonian}
\end{equation}
Here, $c_{{\bm p}\sigma}^\dagger$ is a creation operator of a Fermi atom with pseudospin $\sigma=\uparrow,\downarrow$, describing two atomic hyperfine states. $\xi_{\bm p}=\varepsilon_{\bm p}-\mu=p^2/(2m)-\mu$ is the kinetic energy, measured from the chemical potential $\mu$ (where $m$ is an atomic mass). Since we are considering a two-dimensional system, ${\bm p}=(p_x,p_y)$ is a two-dimensional momentum. $-U (<0)$ is a pairing interaction, which is assumed to be tunable. 
\par
Within LDA, effects of a harmonic trap is conveniently incorporated into the theory by replacing the chemical potential $\mu$ with the LDA expression,
\begin{equation}
\mu(r)=\mu-V(r).
\label{eq.mu}
\end{equation}
Here, $V(r)=m\omega^2_{\rm {tr}}r^2/2$ is a two-dimensional trap potential, where the position $r$ is measured from the trap center, and $\omega_{\rm {tr}}$ is a trap frequency. In LDA, effects of spatial inhomogeneity enter into the theory through the LDA chemical potential $\mu(r)$ in Eq. (\ref{eq.mu}). 
\par
The LDA single-particle thermal Green's function is given by
\begin{equation}
G_{\bm p}(i\omega_n,r)=\frac{1}{G_{\bm p}^0(i\omega_n,r)^{-1}-\Sigma_{\bm p}(i\omega_n,r)},
\label{eq.G}
\end{equation}
where $\omega_n$ is the fermion Matsubara frequency. $G_{\bm p}^0(i\omega_n,r)=[i\omega_n-\xi_{\bm p}(r)]^{-1}$ is the thermal Green's function for a free Fermi gas, where $\xi_{\bm p}(r)=\varepsilon_{\bm p}-\mu(r)$. The LDA self-energy $\Sigma_{\bm p}(i\omega_n,r)$ in Eq. (\ref{eq.G}) describes fluctuation corrections to single-particle excitations. In the ordinary $T$-matrix approximation, it is given by\cite{TSUCHIYA1,TSUCHIYA2,Pietila,PERALI},
\begin{equation}
\Sigma_{\bm p}(i\omega_n,r)=T\sum_{{\bm q},\nu_n}\Gamma_{\bm q}(i\nu_n,r)G_{\bm q-p}^0(i\nu_n-i\omega_n,r),
\label{eq.S}
\end{equation}
where $\nu_n$ is the boson Matsubara frequency. The particle-particle scattering matrix $\Gamma_{\bm q}(i\nu_n,r)$ in Eq. (\ref{eq.S}) describes fluctuations in the Cooper channel, having the form\cite{TSUCHIYA1,TSUCHIYA2,Pietila,PERALI},
\begin{equation}
\Gamma_{\bm q}(i\nu_n,r)=\frac{-U}{1-U\Pi_{\bm q}(i\nu_n,r)}.
\label{Gamma}
\end{equation}
Here,
\begin{equation}
\Pi_{\bm q}(i\nu_n,r)=T\sum_{{\bm{p}},\omega_n}G_{\bm p+q/2}^0(i\nu_n+i\omega_n,r)G_{\bm -p+q/2}^0(-i\omega_n,r)
\end{equation}
is the lowest-order pair-propagator in LDA\cite{NSR,SADEMELO,RANDERIA_BEC}.
\par
In the case of a {\it uniform} two-dimensional Fermi gas, it is well known\cite{Mermin,Hohenberg} that the ordinary BCS-type superfluid phase transition is completely suppressed by strong pairing fluctuations. Instead, the superfluid instability is dominated by the BKT transition\cite{Berezinskii,Kosterlitz1,Kosterlitz2}. On the other hand, in the presence of a harmonic trap, we will show that the BCS-type superfluid phase transition revives (at least within LDA), which is just the same as the case of BEC in a trapped two-dimensional Bose gas\cite{Pethick}. In this case, the LDA superfluid phase transition temperature $T_{\rm c}$ is simply determined from the BCS-type $T_{\rm c}$-equation in the trap center\cite{TSUCHIYA3},
\begin{equation}
1=U\sum_{\bm p}\frac{1}{2\xi_p(r=0)}\tanh\frac{\xi_p(r=0)}{2T_{\rm c}},
\label{eq.gap}
\end{equation} 
together with the equation for the total number $N$ of Fermi atoms,
\begin{equation}
N=\int d{\bm r}n(r).
\label{NUMeq}
\end{equation}
In Eq. (\ref{NUMeq}), $n(r)$ is the LDA particle density, given by
\begin{equation}
n(r)=2T\sum_{{\bm p},\omega_n}G_{\bm p}(i\omega_n,r)e^{i\omega_n\delta},
\end{equation}
where $\delta=+0$ is an infinitesimally small positive number.
\par
As usual, one needs to regularize the $T_{\rm c}$-equation (\ref{eq.gap}) to eliminate the ultraviolet divergence. This can be achieved by introducing the two-dimensional $s$-wave scattering length $a_s$, which is related to the pairing interaction $U$ as\cite{Pietila,Morgan}
\begin{equation}
-U^{-1}=\frac{m}{2\pi}\ln\frac{2}{C\sqrt{-2mE}a_s}-\sum_{\bm p}\frac{1}{E+i\delta-2\varepsilon_p},
\end{equation}
where $C=1.78$, and $E$ is an infinitesimally small energy. In this regularization scheme, the interaction strength is conveniently measured in terms of the binding energy $E_{\rm b}$ of a two-body bound state\cite{Miyake},
\begin{equation}
E_{\rm b}=\frac{2}{C^2ma_{\rm s}^2}.
\label{bind}
\end{equation}
In this scale, the increase of $E_{\rm b}$ corresponds to the increase of the interaction strength. We briefly note that $E_{\rm b}$ always exists in the two-dimensional case, irrespective of the interaction strength $U$\cite{Randeria,Miyake}. This is quite different from the three-dimensional case, where a threshold coupling strength exists to obtain a two-body bound state.
\par

\begin{figure}[t]
\includegraphics[width=0.9\textwidth]{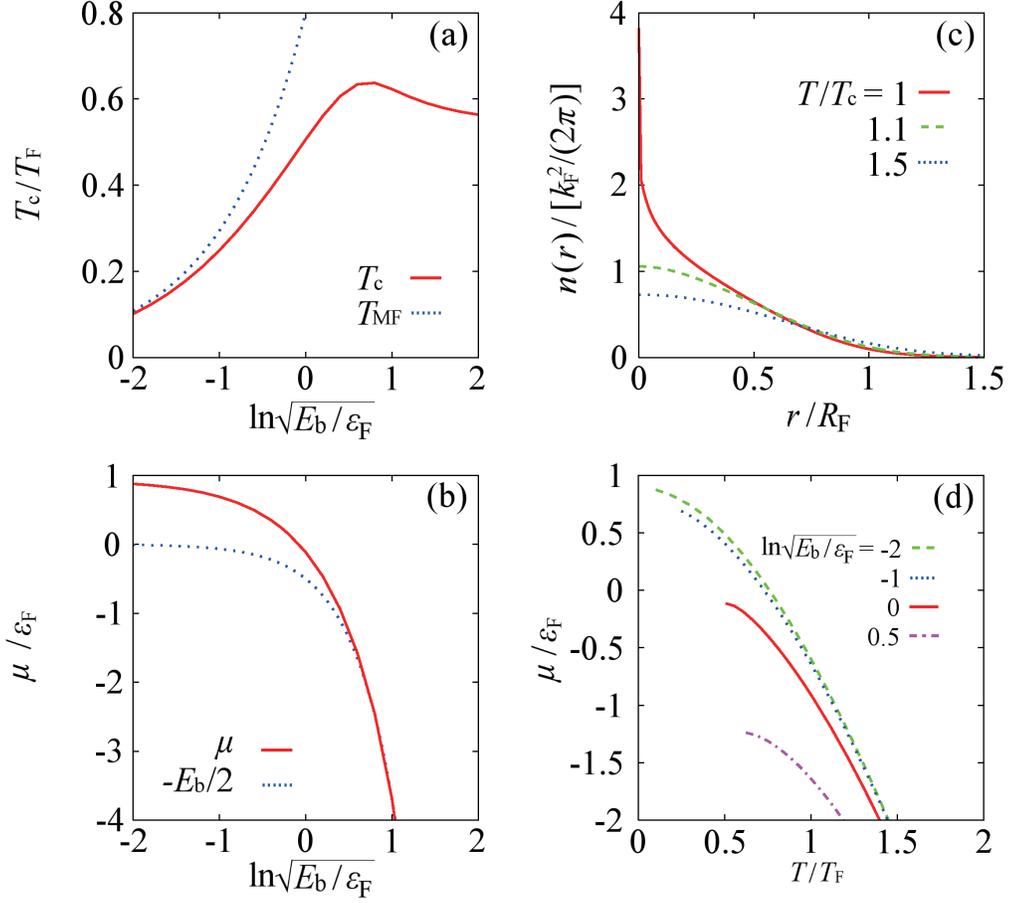}
\caption{(Color online) Calculated $T_{\rm c}$ (a), and $\mu(T=T_{\rm c})$ (b), as functions of the interaction strength, measured in terms of $\ln\sqrt{E_{\rm b}/\varepsilon_{\rm F}}$ (where $E_{\rm b}$ is the molecular binding energy in Eq. (\ref{bind}), and $\varepsilon_{\rm F}=\sqrt{N}\omega_{\rm {tr}}$ is the LDA Fermi energy). In panel (a), $T_{\rm MF}$ is the mean field result. (c) LDA atomic density profile $n(r)$, when $\ln{\sqrt{E_{\rm b}/\varepsilon_{\rm F}}}=0$. $R_{\rm F}=\sqrt{2\varepsilon_{\rm F}/(m\omega_{\rm tr}^2)}$ is the Thomas-Fermi radius, which gives the size of a free Fermi gas at $T=0$ in LDA. (d) Calculated Fermi chemical potential $\mu$ above $T_{\rm c}$.}
\label{fig1}
\end{figure}

Figures \ref{fig1}(a) and (b) show the self-consistent solutions of the coupled Eqs.(\ref{eq.gap}) and (\ref{NUMeq}). As in the three-dimensional case\cite{NSR,SADEMELO,RANDERIA_BEC}, $T_{\rm c}$ and $\mu$ exhibit typical BCS-BEC crossover behaviors in the present two-dimensional case. In panel (a), $T_{\rm c}$ gradually deviates from the mean-field result ($T_{\rm MF}$) with increasing the interaction strength, to approach a constant value in the strong-coupling BEC regime ($\ln\sqrt{E_{\rm b}/\varepsilon_{\rm F}}\gesim 1$, where $\varepsilon_{\rm F}=\sqrt{N}\omega_{\rm {tr}}$ is the LDA Fermi energy). In the BEC limit, the system is well described by a trapped ideal Bose gas of $N/2$ molecules, so that $T_{\rm c}$ is determined from the LDA molecular number equation\cite{Pethick},
\begin{equation}
{N \over 2}=\int d{\bf r}\sum_{\bm q}
{1 \over e^{[q^2/(4m)+2V(r)]/T_{\rm c}}-1},
\end{equation}
which gives
\begin{equation}
T_{\rm c}={\sqrt{3} \over \pi}T_{\rm F}=0.551T_{\rm F}.
\end{equation}
Here, $T_{\rm F}$ is the Fermi temperature. Figure \ref{fig1}(a) indicates that the crossover from the BCS regime to the BEC regime occurs in the region $-1\lesssim\ln\sqrt{E_{\rm b}/\varepsilon_{\rm F}}\lesssim 1$.
\par
In Fig.\ref{fig1}(b), the Fermi chemical potential $\mu$ gradually deviates from the Fermi energy $\varepsilon_{\rm F}$ with increasing the interaction strength, to be negative in the strong-coupling BEC regime. In the BEC regime, $2|\mu|$ approaches the molecular binding energy $E_{\rm b}$, reflecting that the system becomes a gas of tightly bound molecules having this binding energy. 
\par  
We note that, although our LDA calculation gives a finite $T_{\rm c}$ in the presence of a trap, one also sees a sign of the vanishing BCS superfluid phase transition in the uniform case. As shown in Fig.\ref{fig1}(c), $n(r)$ has a peak structure at $T_{\rm c}$, which is known as an artifact of LDA\cite{Pethick}. While the peak value is finite in the three-dimensional case, $n(r=0)$ diverges in the two-dimensional case. Indeed, in the strong-coupling BEC limit, one finds,
\begin{equation}
n(r)=-\frac{mT_{\rm c}}{\pi}\ln{\left(1-e^{-\frac{2V(r)}{T_{\rm c}}}\right)},
\end{equation} 
which behaves as $n(r)\sim-\ln(r)$ around $r=0$. However, since the total number equation (\ref{NUMeq}) converges after taking the radial integration ($\int dr rn(r)$) of this singular particle density $n(r)$, one obtains self-consistent solutions of the coupling equations (\ref{eq.gap}) and (\ref{NUMeq}). In the uniform case, on the other hand, the divergence of the (uniform) particle density occurs everywhere, when the $T_{\rm c}$-equation (\ref{eq.gap}) is satisfied. As a result, the equation for the total number $N$ of Fermi atoms also diverges. Because of this singularity, the coupled equations (\ref{eq.gap}) and (\ref{NUMeq}) do not give any self-consistent solution in the uniform case, leading to the vanishing BCS phase transition. According to the Thouless criterion\cite{Thouless}, the $T_{\rm c}$-equation (\ref{eq.gap}) is directly related to the pole of the particle-particle scattering matrix $\Gamma_{\bm q}(i\nu_n)$ in Eq. (\ref{Gamma}) at $q=\nu_n=0$. Since this quantity physically describes pairing fluctuations, the vanishing $T_{\rm c}$ is due to strong pairing fluctuations in a two-dimensional Fermi gas\cite{Mermin,Hohenberg}.
\par
Since all the current experiments on two-dimensional Fermi gases have been done in the normal state\cite{KOHL,FROHLICH,SOMMER}, we also consider the region above $T_{\rm c}$ in this paper. In the normal state, one may only solve the number equation (\ref{NUMeq}) to obtain $\mu$ shown in Fig.\ref{fig1}(d). Then, the LDA single-particle spectral weight $A_{\bm p}(\omega,r)$, as well as the local density of states $\rho(\omega,r)$, are conveniently calculated from the analytic continued Green's function as,
\begin{equation}
A_{\bm p}(\omega,r)=-\frac{1}{\pi}{\rm Im}[G_{\bm p}(i\omega_n\to\omega+i\delta,r)],
\label{LSW}
\end{equation}
\begin{equation}
\rho(\omega,r)=\sum_{\bm p}A_{\bm p}(\omega,r).
\label{LDOS}
\end{equation}
\par
The photoemission spectrum is also related to the spectral weight $A_{\bm p}(\omega,r)$ in Eq. (\ref{LSW}). In this experiment, atoms in one of the two hyperfine states ($\equiv|\uparrow\rangle$) is transfered to another hyperfine state $|3\rangle$ by rf-pulse, and the tunneling current between these states is measured. Since the final state interaction can be safely ignored in a $^{40}$K Fermi gas\cite{STEWART,GAEBLER}, one may safely treat $|3\rangle$ as a non-interacting state. In this case, the LDA local photoemission current $I({\bm p},\Omega,r)$ is given by\cite{TSUCHIYA3}
\begin{equation}
I({\bm p},\Omega,r)=2\pi t_{\rm F}^2A_{\bm p}(\xi_{\bm p}(r)-\Omega,r)f(\xi_{\bm p}(r)-\Omega),
\label{eq.100}
\end{equation}
where $f(z)$ is the Fermi distribution function. $t_{\rm F}$ is a transfer matrix element between $|\uparrow\rangle$ and $|3\rangle$. $\Omega=\omega_{\rm L}-\omega_3$ is the energy difference between the incident photon energy $\omega_{\rm L}$ and the energy $\omega_3$ of the final state $|3\rangle$. Since the current photoemission-type experiments\cite{KOHL,FROHLICH,SOMMER,STEWART,GAEBLER} do not have enough spatial resolution, one needs to take the spatial average of Eq. (\ref{eq.100}) over the gas cloud, which gives
\begin{equation}
I_{\rm ave}({\bm p},\Omega)=\frac{2\pi t_{\rm F}^2}{\pi R_{\rm F}^2}\int d{\bm r}A_{\bm p}(\xi_{\bm p}(r)-\Omega,r)f(\xi_{\bm p}(r)-\Omega),
\label{eq.101}
\end{equation}
where $R_{\rm F}=\sqrt{2\varepsilon_{\rm F}/(m\omega_{\rm tr}^2)}$ is the Thomas-Fermi radius. The observed photoemission spectrum\cite{KOHL} ($\equiv \overline{A_{\bm p}(\omega)f(\omega)}$) is then given by\cite{TSUCHIYA3} 
\begin{equation}
\overline{A_{\bm p}(\omega)f(\omega)}=
I_{\rm ave}({\bm p},\Omega\to \xi_p-\omega).
\label{eq.102}
\end{equation}
In later sections, we numerically evaluate $\rho(\omega,r)$ in Eq. (\ref{LDOS}), as well as $\overline{A_{\bm p}(\omega)f(\omega)}$ in Eq. (\ref{eq.102}), to examine strong-coupling effects on single-particle properties of a two-dimensional Fermi gas.
\par
Before ending this section, we briefly explain our idea about how to compare our theoretical results with experimental data. Although the prediction for the value of $T_{\rm c}$ is usually difficult in a Fermi superfluid, it is known in the field of superconductivity that various superconducting phenomena can be well explained theoretically, when one uses the {\it scaled} temperature $T/T_{\rm c}$ (even if the theory cannot reproduce the observed value of $T_{\rm c}$). However, the superfluid phase transition has not been realized yet in a two-dimensional Fermi gas. Thus, apart from quantitative discussions, it seems difficult to quantitatively compare the calculated temperature dependence of a physical quantity with experiment data in the current stage of research. However, we point out that the (pseudo)gap size $\omega_0$ in the photoemission spectrum at $p=0$ observed in a two-dimensional $^{40}$K Fermi gas is almost $T$-independent in the wide temperature region, $0.27T_{\rm F}\le T\le 0.65T_{\rm F}$\cite{KOHL}. Thus, as an alternative to using the scaled temperature $T/T_{\rm c}$, $\omega_0$ would be useful for the comparison of theory with experiment in the current stage of research. In Sec. IV, we will take this strategy to assess the pseudogap scenario.
\par
\begin{figure}[t]
\includegraphics[width=0.8\textwidth]{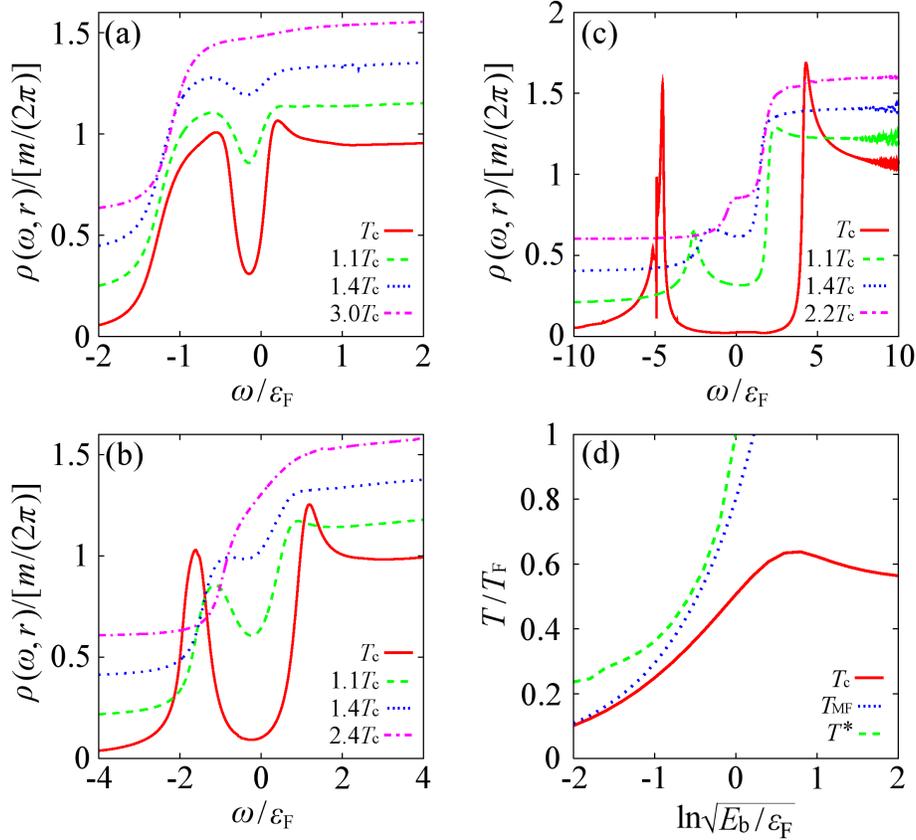}
\caption{(Color online) Calculated local density of states $\rho(\omega,r)$ at $r=0.01R_{\rm F}$. (a) $\ln{\sqrt{E_{\rm b}/\varepsilon_{\rm F}}}=-2$. (b) $\ln{\sqrt{E_{\rm b}/\varepsilon_{\rm F}}}=-1$. (c) $\ln{\sqrt{E_{\rm b}/\varepsilon_{\rm F}}}=0$. In these panels, as well as in Fig.\ref{fig3}, we have offset the results by 0.2. (d) Pseudogap temperature $T^*$, determined from the temperature dependence of the local density of states $\rho(\omega,r)$. Fine structures seen in panel (c) are due to a computational problem in calculating the analytic continued Green's function.
}
\label{fig2}
\end{figure}

\begin{figure}[t]
\includegraphics[width=0.5\textwidth]{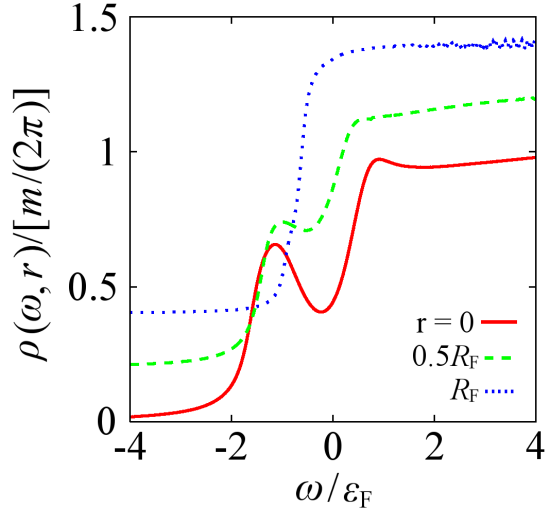}
\caption{(Color online) Spatial variation of the local density of states $\rho(\omega,r)$ at $T=1.1T_{\rm c}$. We set $\ln{\sqrt{E_{\rm b}/\varepsilon_{\rm F}}}=-1$. 
}
\label{fig3}
\end{figure}

\section{Local density of states and inhomogeneous pseudogap phenomenon}
\par
Figures \ref{fig2}(a)-(c) show the local density of states $\rho(\omega,r)$ at $r=0.01R_{\rm F}$ in the BCS-BEC crossover regime of a two-dimensional Fermi gas\cite{note}. We find that $\rho(\omega,r)$ has a gap-like (pseudogap) structure near $T_{\rm c}$. In particular, when $\ln\sqrt{E_{\rm b}/\varepsilon_{\rm F}}=0$ (panel (c)), the pseudogap structure at $T_{\rm c}$ is very similar to the BCS-type superfluid density of states with sharp coherence peaks at the gas edges. Since the superfluid order parameter vanishes in these figures, the pseudogap in $\rho(\omega,r)$ originates from two-dimensional pairing fluctuations. Indeed, noting that the particle-particle scattering matrix $\Gamma_{{\bm q}=0}(i\nu_n=0,r=0)$ (which physically describes pairing fluctuations) diverges at $T_{\rm c}$, we may approximate Eq. (\ref{eq.S}) to $\Sigma_{\bm p}(i\omega_n,r=0)\simeq -G_{\bm -p}^0(-i\omega_n,r=0)\Delta_{\rm pg}^2$ near $T_{\rm c}$, where $\Delta_{\rm pg}^2=-T\sum_{{\bm q},\nu_n}\Gamma_{\bm q}(i\nu_n,r=0)$ is sometimes referred to as the pseudogap parameter in the literature\cite{LEVIN,TSUCHIYA1}. Substituting this into Eq.(\ref{eq.G}), we obtain
\begin{eqnarray}
G_{\bm p}(i\omega_n,r=0)
&=&
{1 \over \displaystyle i\omega_n-\xi_{\bm p}-{\Delta_{\rm pg}^2 \over i\omega_n+\xi_{\bm p}}}
\nonumber
\\
&=&
-{i\omega_n+\xi_{\bm p} \over \omega_n^2+\xi_{\bm p}^2+\Delta_{\rm pg}^2},
\label{eq.BCS}
\end{eqnarray}
which is the same form as the mean-field Green's function in the BCS state, 
\begin{eqnarray}
G^{\rm BCS}_{\bm p}(i\omega_n)
=-{i\omega_n+\xi_{\bm p} \over \omega_n^2+\xi_{\bm p}^2+\Delta^2}.
\label{eq.BCS2}
\end{eqnarray}
Thus, pairing fluctuations (that are described by $\Delta_{\rm pg}$ in Eq. (\ref{eq.BCS})) are found to work like the BCS gap parameter $\Delta$, leading to the pseudogap in the single-particle density of states, as shown in Fig.\ref{fig2}(a)-(c). We also find from the first line in Eq. (\ref{eq.BCS}) that pairing fluctuations ($\Delta_{\rm pg}$) couple the particle branch $\omega_{\rm p}\equiv\xi_{\bm p}$ with the hole branch $\omega_{\rm h}\equiv-\xi_{\bm p}$, leading to the level repulsion between the two around $p=\sqrt{2m\mu}$ (when $\mu\ge 0$). In this sense, the pseudogap may be viewed as a particle-hole coupling phenomenon, induced by strong pairing fluctuations.
\par
When $\ln\sqrt{E_{\rm b}/\varepsilon_{\rm F}}=-2$, Fig.\ref{fig1}(a) shows that $T_{\rm c}$ is almost equal to the mean-field value ($T_{\rm MF}$). In addition, the Fermi chemical potential $\mu(T=T_{\rm c})$ is also close to the Fermi energy $\varepsilon_{\rm F}$ there, as shown in Fig.\ref{fig1}(b). However, even in this weak-coupling case, $\rho(\omega,r)$ exhibits a pseudogap around the trap center, as shown in Fig.\ref{fig2}(a). Since such a clear pseudogap does not appear in the weak-coupling regime of a three-dimensional Fermi gas\cite{WATANABE1}, the two-dimensionality is found to enhance the pseudogap phenomenon, as expected.
\par
In Figs. \ref{fig2}(a)-(c), we find that the pseudogap gradually disappears, as one increases the temperature above $T_{\rm c}$. When we conveniently define the pseudogap temperature $T^*$ as the temperature at which the dip structure in $\rho(\omega,r)$ completely disappears in the trap center, Fig.\ref{fig2}(d) shows that $T^*$ is relatively high even in the weak-coupling regime ($\ln\sqrt{E_{\rm b}/\varepsilon_{\rm F}}\lesssim -1$). We briefly note that, in the three-dimensional case\cite{TSUCHIYA3,WATANABE2}, the pseudogap temperature is only slightly higher than $T_{\rm c}$ in this regime. Thus, the present result also indicates the enhancement of the pseudogap phenomenon by the two-dimensionality of the system. 
\par
\begin{figure}[t]
\includegraphics[width=.4\textwidth]{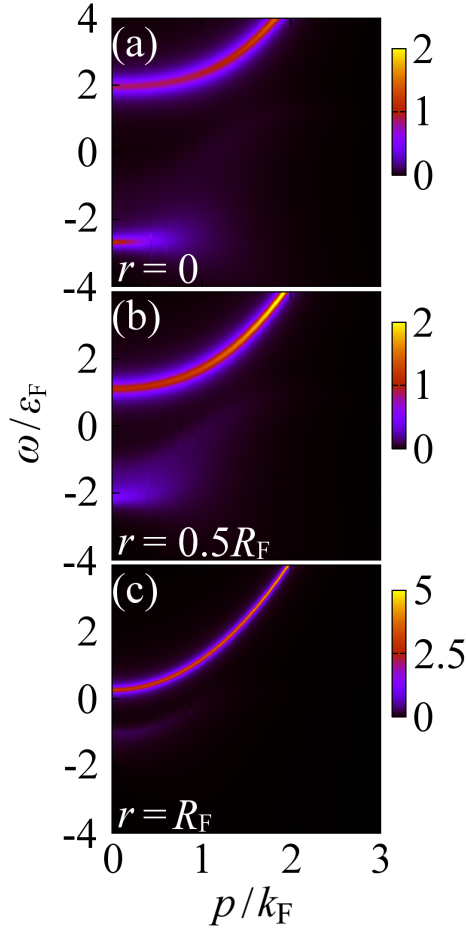}
\caption{(Color online) Calculated intensity of single-particle spectral weight $A_{\bm p}(\omega,r)$ at $T=1.1T_{\rm c}$. We take $\ln[E_b/\varepsilon_F]=0$. The intensity is normalized by $\varepsilon_{\rm F}^{-1}$.}
\label{fig4}
\end{figure}

\begin{figure}[t]
\includegraphics[width=.4\textwidth]{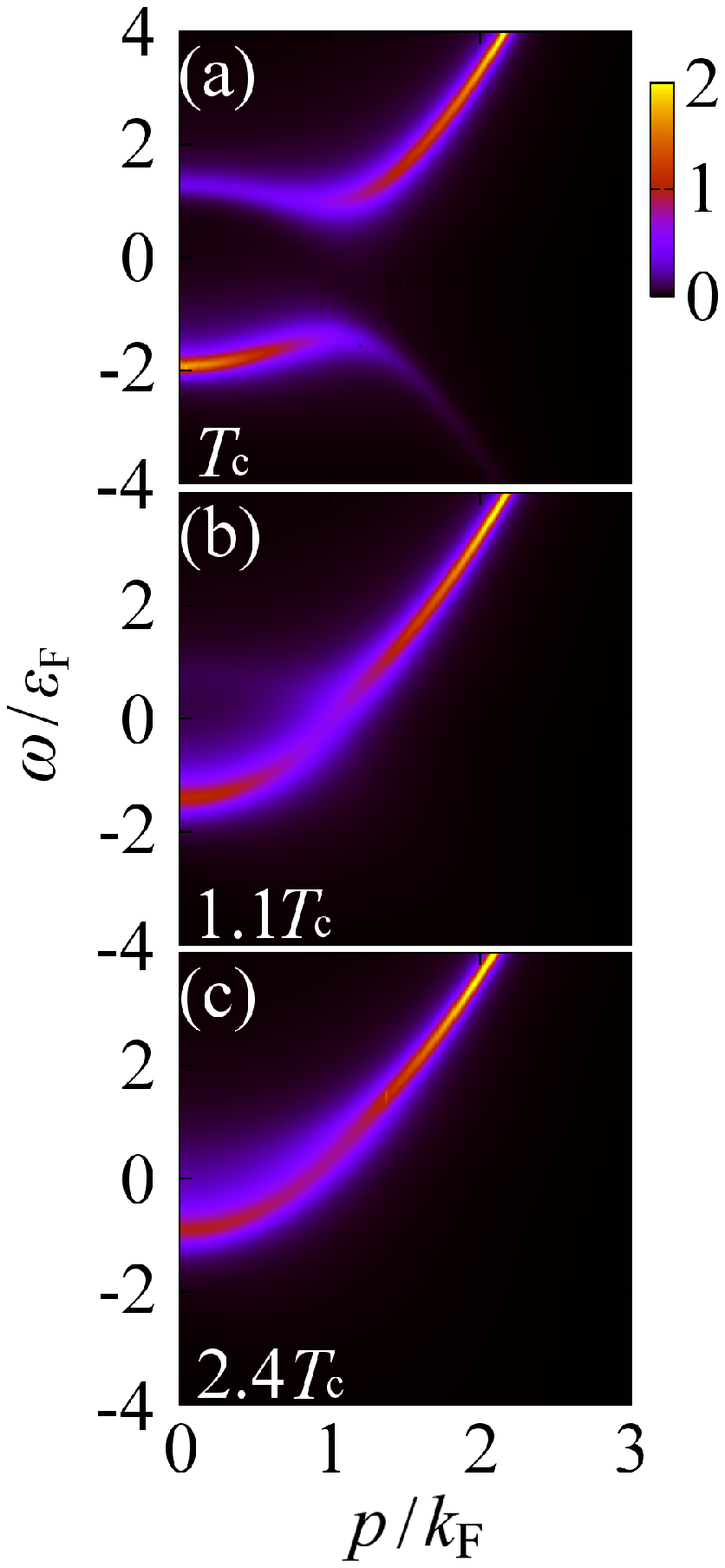}
\caption{(Color online) Calculated intensity of single-particle spectral weight $A_{\bm p}(\omega,r)$ at $r=0.01R_{\rm F}$. We take $\ln[E_{\rm b}/\varepsilon_F]=-1$. The intensity is normalized by $\varepsilon_{\rm F}^{-1}$.}
\label{fig5}
\end{figure}

Figure \ref{fig3} shows the spatial variation of the local density of states. As in the three-dimensional case\cite{TSUCHIYA3}, the pseudogap remarkably depends on the spatial position. That is, the dip structure gradually becomes obscure, as one goes away from the trap center. Around the edge of the gas cloud ($r\sim R_{\rm F}$), $\rho(\omega,r)$ is almost equal to the density of states $\rho_{\rm 2D}(\omega)$ in a two-dimensional free Fermi gas,
\begin{equation}
\rho_{\rm 2D}(\omega)={m \over 2\pi}\Theta(\omega-\mu(r\sim R_{\rm F})),
\end{equation}
where $\Theta(x)$ is the step function. Thus, the pseudogapped density of states and the free-particle-like density of states coexist in a trapped Fermi gas (although there is no clear boundary between them).
\par
We briefly note that one can also see the pseudogap phenomenon in the spectral weight $A_{\bm p}(\omega,r)$. In Fig.\ref{fig4}, while $A_{\bm p}(\omega,r)$ exhibits a double peak (pseudogap) structure near the trap center (panel (a)), the lower peak in the spectral weight gradually becomes obscure, as one approaches the edge of the gas cloud (panels (b) and (c).) At $r=R_{\rm F}$ (panel (c)), $A_{\bm p}(\omega,r)$ is dominated by a single peak line, which is close to the free particle dispersion ($\omega\simeq p^2/(2m)-\mu(R_{\rm F})\simeq p^2/(2m)$). This spatial dependence of $A_{\bm p}(\omega,r)$ is consistent with that of the local density of states $\rho(\omega,r)$ shown in Fig.\ref{fig3}. In addition, as expected from the temperature dependence of the pseudogap structure shown in Figs.\ref{fig2}(a)-(c), Fig.\ref{fig5} shows that the double peak structure in the spectral weight $A_{\bm p}(\omega,r\simeq 0)$ gradually disappears, as one increases the temperature from $T_{\rm c}$. Since the spectral weight is deeply related to the photoemission spectrum (See Eq. (\ref{eq.101}).), the above results make us expect that the latter is a useful quantity to observe the pseudogap phenomenon in a two-dimensional Fermi gas, which we will confirm in the next section.
\par
\begin{figure}[t]
\includegraphics[width=0.5\textwidth]{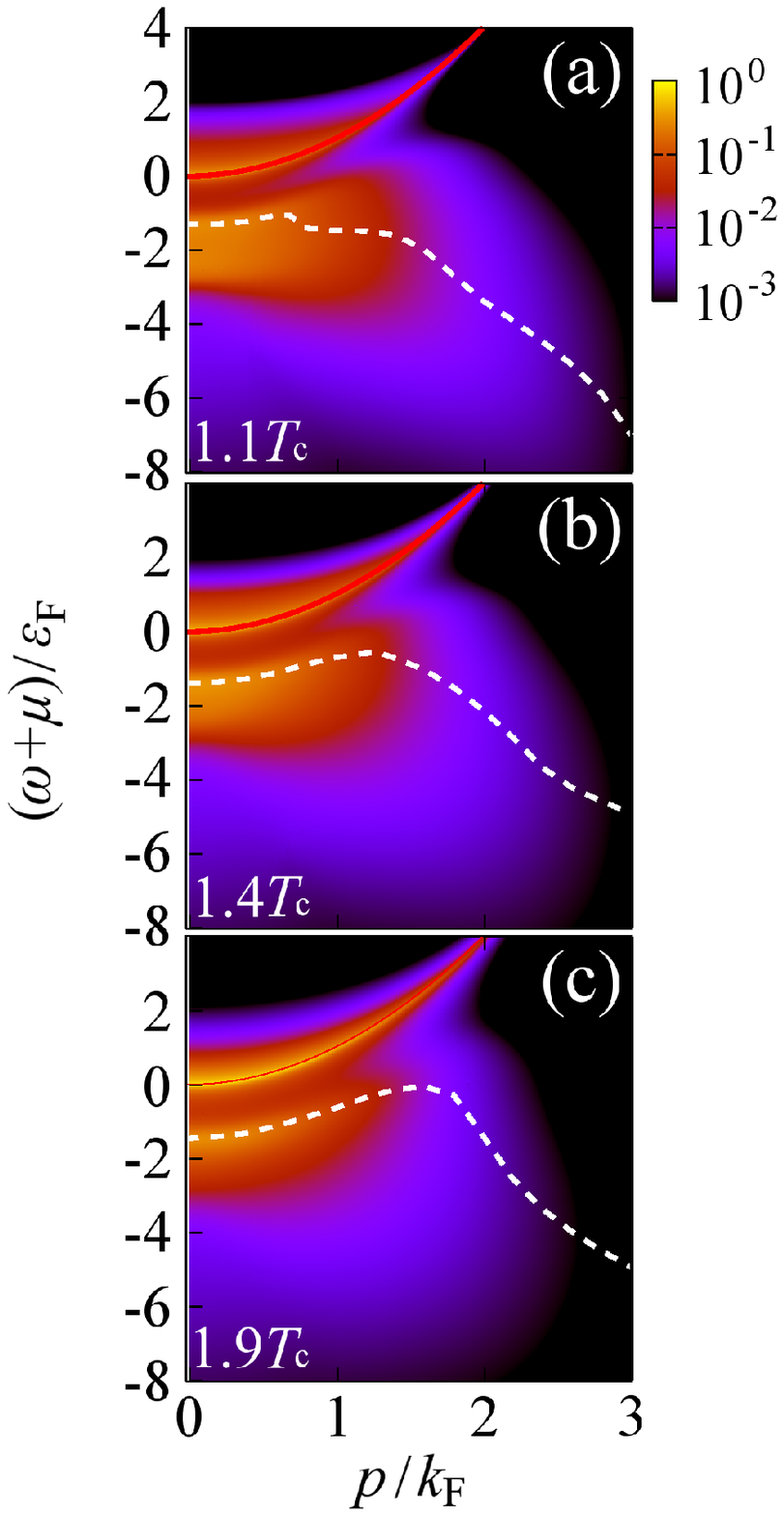}
\caption{(Color online) Calculated intensity of photoemission spectrum $\overline{A_{\bm p}(\omega)f(\omega)}$ in the normal state. We take $\ln{\sqrt{E_{\rm b}/\varepsilon_{\rm F}}}=0$. The spectral intensity is normalized by $2\pi t_{\rm F}^2/\varepsilon_{\rm F}$. In each panel, the solid line shows the free particle dispersion, $\omega+\mu=p^2/(2m)$, which is close to the upper peak line of the spectrum. The dashed line shows the lower peak line of the spectrum.}
\label{fig6}
\end{figure}

\begin{figure}[t]
\includegraphics[width=.9\textwidth]{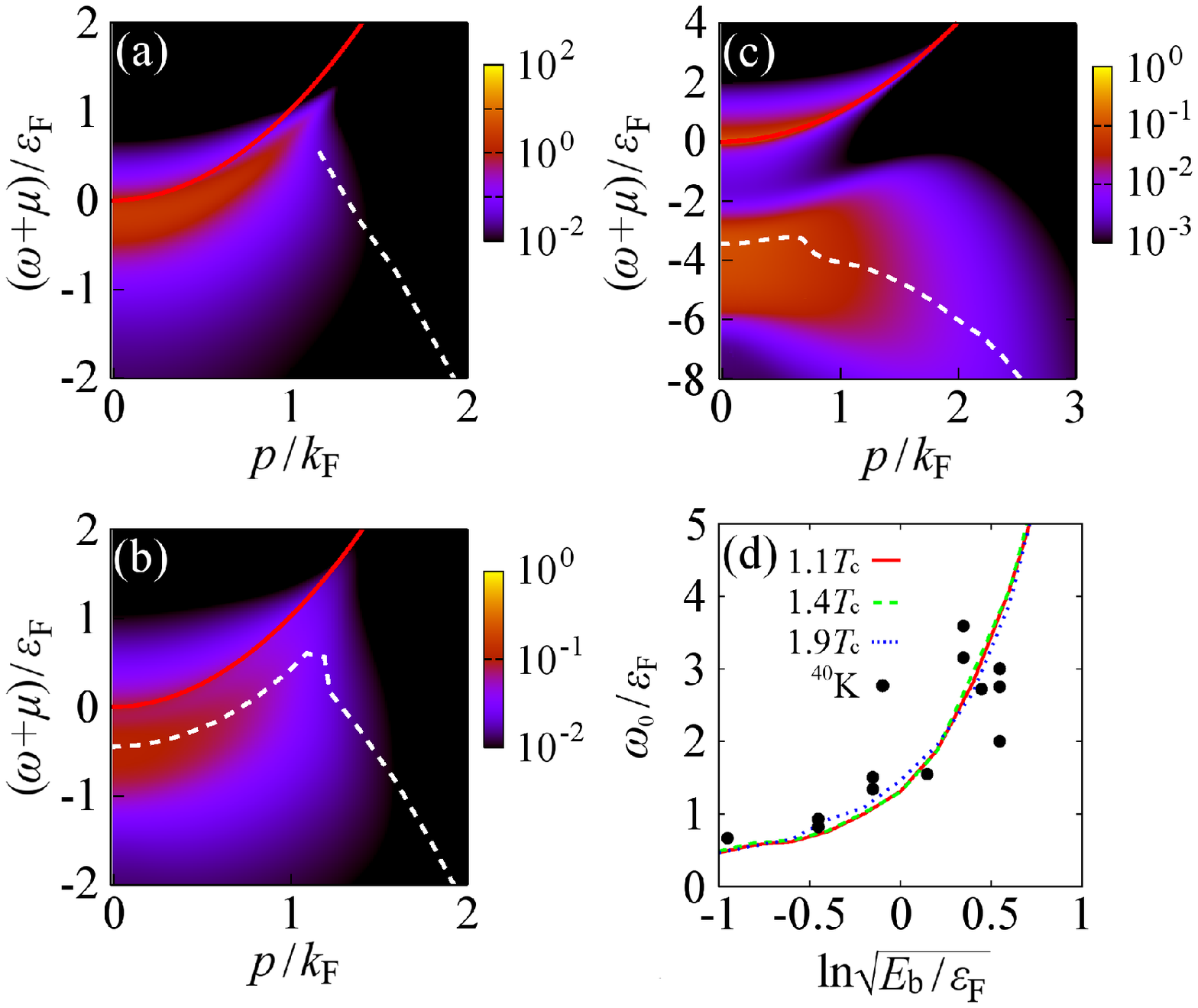}
\caption{(Color online) Calculated intensity of photoemission spectrum $\overline{A_{\bm p}(\omega)f(\omega)}$ at $T=1.1T_{\rm c}$. (a) $\ln{\sqrt{E_{\rm b}/\varepsilon_{\rm F}}}=-2$. (b) $\ln{\sqrt{E_{\rm b}/\varepsilon_{\rm F}}}=-1$. (c) $\ln{\sqrt{E_{\rm b}/\varepsilon_{\rm F}}}=0.5$. Panel (d) shows the peak-to-peak energy $\omega_0$ of the calculated photoemission spectrum at $p=0$. Solid circles are experimental data for a $^{40}$K Fermi gas\cite{KOHL}. The dashed line shows the lower peak line of the spectrum.}
\label{fig7}
\end{figure}

\section{Photoemission spectrum in a two-dimensional Fermi gas}
\par
Figure \ref{fig6} shows the intensity of the photoemission spectrum $\overline{A_{\bm p}(\omega)f(\omega)}$ in a two-dimensional Fermi gas. In the intermediate coupling case ($\ln\sqrt{E_{\rm b}/\varepsilon_{\rm F}}=0$), one sees a clear double peak structure, as observed in a $^{40}$K Fermi gas\cite{KOHL}. In the case of a free Fermi gas, a single peak line along the free-particle dispersion $\omega+\mu=p^2/(2m)$ only appears in the spectrum. In addition, the superfluid order parameter vanishes in Fig.\ref{fig6}. Thus, the gap-like (double peak) structure in Fig.\ref{fig6} is considered to originate from two-dimensional pairing fluctuations. 
\par
In Fig.\ref{fig6}, the upper peak line is well described by the free particle dispersion, $\omega+\mu=p^2/(2m)$. This mean that this branch is dominated by single-particle excitations in the outer region of the gas cloud ($r\sim R_{\rm F}$) where a free-particle-like local density of states shown in Fig.\ref{fig3} is obtained. Actually, since the inner region of the gas cloud also contributes to this branch to some extent, it has a finite spectral width.
\par
On the other hand, the lower peak line in Fig.\ref{fig6} (which does not appear in a free Fermi gas) comes from the pseudogapped single-particle excitations around the trap center. Since the pseudogap phenomenon is spatially inhomogeneous in the presence of a trap (See Figs.\ref{fig3} and \ref{fig4}.), the spatial average of the local photoemission spectrum makes this branch broad, as shown in Fig.\ref{fig6}. 
\par
We note that the lower branch in Fig.\ref{fig6} exhibits the so-called back-bending behavior. With increasing the temperature, the momentum region where this branch shows an upward behavior becomes wide. In addition, the energy difference ($\equiv\omega_0$) between the two peak lines at $p=0$ is found to be insensitive to the temperature, at least up to $T\sim 2T_{\rm c}$. These behaviors are consistent with the recent experiment on a $^{40}$K Fermi gas\cite{KOHL}. As mentioned in Sec. II, the latter property is useful for the comparison of our theory with experimental data, which we will use soon later. 
\par
Figure \ref{fig7}(a)-(c) show that, although the pseudogap structure (double peak structure) is clearly seen in the photoemission spectrum in the BEC side (panel (c)), it gradually becomes obscure, as one approaches the weak-coupling BCS regime. The double peak structure still remains when $\ln\sqrt{E_{\rm b}/\varepsilon_{\rm F}}=-1$ (Fig.\ref{fig7}(b)), a broad spectral peak only exists around $p=0$ when $\ln\sqrt{E_{\rm b}/\varepsilon_{\rm F}}=-2$ (Fig.\ref{fig7}(a)). In the latter case, the local density of states $\rho(\omega,r\sim 0)$ has a clear pseudogap structure near $T_{\rm c}$, as shown in Fig.\ref{fig2}(a). Thus, the disappearance of the double peak structure in the low momentum regime of the photoemission spectrum is a result of the spatial average of the inhomogeneous pseudogap.
\par
Figure \ref{fig7}(d) shows the peak-to-peak energy at $p=0$ ($\omega_0$) in the photoemission spectrum $\overline{A_{\bm p}(\omega)f(\omega)}$. As expected from Fig.\ref{fig6}, $\omega_0$ is almost $T$-independent up to $T\sim 2T_{\rm c}$ in the whole BCS-BEC crossover region. In addition, our result agrees well with the recent experiment on a $^{40}$K Fermi gas\cite{KOHL} (solid circles in Fig.\ref{fig7}(d)). Here, we emphasize that no fitting parameter is introduced in obtaining this result. Since the double peak structure in our theoretical results originates from the pseudogap phenomenon associated with two-dimensional pairing fluctuations, this agreement supports the pseudogap scenario as the mechanism of the double peak structure observed in the recent photoemission-type experiment on a two-dimensional $^{40}$K Fermi gas\cite{KOHL}.
\par
\section{Summary}
To summarize, we have discussed single-particle properties and strong-coupling effects in the BCS-BEC crossover regime of a two-dimensional Fermi gas. Within the framework of a combined $T$-matrix theory with LDA, we have calculated the local density of states $\rho(\omega,r)$ in the normal state above $T_{\rm c}$. We showed that $\rho(\omega,r)$ has a pseudogap near $T_{\rm c}$, originating from strong two-dimensional pairing fluctuations. In the intermediate coupling regime ($\ln\sqrt{E_{\rm b}/\varepsilon_{\rm F}}=0$), the pseudogap structure at $T_{\rm c}$ is very similar to the BCS superfluid density of states, being accompanied by sharp coherence peaks at gap edges (although the superfluid order parameter vanishes at $T_{\rm c}$). Even in the weak-coupling BCS regime ($\ln\sqrt{E_{\rm b}/\varepsilon_{\rm F}}=-2$), where both $T_{\rm c}$ and $\mu$ are well described by the simple mean-field theory, a dip structure is still obtained. We also showed that the pseudogap remains to a high temperature, in comparison with the three-dimensional case. These results indicate that the pseudogap phenomenon is enhanced by two-dimensional pairing fluctuations. 
\par
In the presence of a trap potential, the pseudogap becomes spatially inhomogeneous. That is, while the pseudogap dominates single-particle excitations in the trap center, the local density of states near the edge of the gas cloud is still similar to that for a free Fermi gas. Including this inhomogeneous pseudogap effect, we calculated the photoemission spectrum. Although the spatial average of the spectrum smears the pseudogap structure to some extent, the calculated spectrum still exhibits a double peak structure, when $\ln\sqrt{E_{\rm b}/\varepsilon_{\rm F}}\gesim -1$. The peak-to-peak energy $\omega_0$ at $p=0$ is almost $T$-independent in the wide temperature region ($T_{\rm c}\le T\lesssim 2T_{\rm c}$), which is consistent with the recent experiment on a $^{40}$K Fermi gas\cite{KOHL}. In addition, the value of $\omega_0$ agrees well with this experiment over the entire BCS-BEC crossover region. This agreement indicates that the anomalous double peak structure observed in the photoemission spectrum\cite{KOHL} may be understood as a pseudogap phenomenon, originating from two-dimensional pairing fluctuations. Since the pseudogap is a fundamental phenomenon in a strongly interacting Fermi system, our results would be useful for the study how this many-body phenomenon is affected by the dimensionality of a system. 
\par
In the current stage of research for a two-dimensional Fermi gas, the achievement of the superfluid phase transition is one of the most important challenges. While the combined $T$-matrix theory with LDA used in this paper gives a finite BCS phase transition temperature, the BKT transition is predicted in a uniform Fermi gas\cite{Berezinskii,Kosterlitz1,Kosterlitz2,Randeria,Melo1,Melo2,Zhang,Tempere}. Thus, it would be interesting to clarify whether the superfluid phase transition in a trapped two-dimensional Fermi is close to the BCS-type or the BKT-type. For this purpose, an extension of our work to include the BKT transition would be an important future problem. 
\par
In addition, since pairing fluctuations become strong with decreasing the temperature near $T_{\rm c}$, a phenomenon which is sensitive this temperature dependence would be helpful to see to what extend the current experiment approaches the two-dimensional superfluid instability. For this purpose, the peak-to-peak energy $\omega_0$ discussed in this paper is {\it not} useful, because it is insensitive to the temperature (although it is useful for the comparison of theory with experiment in the current stage of research). Thus, the search for a two-dimensional pseudogap phenomenon which has a remarkable temperature dependence near $T_{\rm c}$ would be also another important future challenge. 
\par
\acknowledgments
We would like to thank M. K\"{o}hl for fruitful discussions. We also thank R. Hanai for reading this manuscript and giving useful comments. R.W. was supported by the Japan Society for the Promotion of Science. Y.O. was supported by Grant-in-Aid for Scientific research from MEXT in Japan (No.23500056, No.25400418, No.25105511).
\par



\begin{thebibliography}{99}
\bibitem{note} As shown in Fig.\ref{fig1}(c), the LDA particle density at $r=0$ diverges at $T_{\rm c}$, so that we cannot calculate $\rho(\omega,r=0)$ at this temperature. To avoid this singularity, we have calculated the local density of states slightly away from the trap center. When $T>T_{\rm c}$, such a singularity is absent (See Fig.\ref{fig1}(c).), so that we can calculate $\rho(\omega,r)$ everywhere in the gas cloud.
\end{thebibliography}
\end{document}